\documentclass[preprint,review,12pt]{elsarticle}
\usepackage{graphicx}
\usepackage{epsfig,graphics,graphicx}
\usepackage{ccaption}
\usepackage{amssymb}
\usepackage{amsthm}
\usepackage{amsfonts}

\newcommand{\ba}{\begin{eqnarray}}
\newcommand{\ea}{\end{eqnarray}}

\journal{Nuclear Physics A}

\begin{document}

\begin{frontmatter}

%% Title, authors and addresses

%% use the tnoteref command within \title for footnotes;
%% use the tnotetext command for theassociated footnote;
%% use the fnref command within \author or \address for footnotes;
%% use the fntext command for theassociated footnote;
%% use the corref command within \author for corresponding author footnotes;
%% use the cortext command for theassociated footnote;
%% use the ead command for the email address,
%% and the form \ead[url] for the home page:
%% \title{Title\tnoteref{label1}}
%% \tnotetext[label1]{}
  \author{L. Portugal}
  \ead{licinio@if.ufrj.br}

  \author{T. Kodama}
  \ead{tkodama@if.ufrj.br}  
 \address{Instituto de F\'\i sica, Universidade Federal do Rio de Janeiro, \\
Caixa Postal 68528, Rio de Janeiro, RJ 21941-972, Brazil.}

\title{Gluon Saturation Effects at the Nuclear Surface: 
Inelastic Cross Section of Proton-Nucleus in the
Ultra High Energy Cosmic Ray domain}

\begin{abstract}
Considering the high-energy limit of the QCD gluon distribution inside a
nucleus, we calculate the proton-nucleus total inelastic cross section using a
simplified dipole model. We show that, if gluon saturation occurs in the nuclear surface
region, the total cross section of proton-nucleus collisions increases more
rapidly as a function of the incident energy compared to that of a Glauber-type
estimate. We discuss the implications of this with respect to 
recent ultra-high-energy cosmic ray experiments.

\end{abstract}

\begin{keyword}

Gluon distribution in nuclei, p-A reaction cross section, UHECR

%% PACS codes here, in the form: \PACS code \sep code

%% MSC codes here, in the form: \MSC code \sep code
%% or \MSC[2008] code \sep code (2000 is the default)

\end{keyword}

\end{frontmatter}

\section{Introduction\noindent}

In the ultra-high-energy domain, the mechanism of inelastic hadronic
collisions is dominated by the contribution from small-$x$ gluons. This is
the basic reason why the total or inelastic proton-proton collision cross
section increases as a function of the incident energy. A simple way to see
this is to use the eikonal expression for the reaction cross section in the
impact parameter representation as%
\begin{equation}
\sigma _{r}=\int d^{2}\mathbf{b}[1-\exp (-2\chi (\mathbf{b},s))],
\label{inela}
\end{equation}%
where the eikonal $\chi (\mathbf{b},s)$ counts essentially the total number
of possible scattering centers of the constituents inside the target
\textquotedblleft seen\textquotedblright\ by the projectile passing through
the target along a straight line with the impact parameter $\mathbf{b,}$ and
with center-of-mass energy $\sqrt{s}$.

If the target is thick enough, the eikonal function is much larger than
unity ( $\chi (\mathbf{b},s)\gg 1$) in the central region, but it falls down
to zero in the surface region. The integrand of Eq.(\ref{inela}) keeps a
value almost equal to unity while $\chi (\mathbf{b},s)\gg 1,$ and falls down
quickly to zero near the surface. From this we may determine an effective
radius $b_{1/2}$ such that the integral (\ref{inela}) is approximately given
by 
\begin{equation}
\sigma _{r}\simeq \pi b_{1/2}^{2}.
\end{equation}%
Here $b_{1/2}$ might be estimated by fixing the value of $\chi (\mathbf{b}%
_{1/2},s)$, for example, as 
\begin{equation}
\chi (\mathbf{b}_{1/2},s)=\ln 2,
\end{equation}%
so that the effective radius depends on the energy $b_{1/2}=b_{1/2}\left( 
\sqrt{s}\right) $. Let us assume that the eikonal function factorizes in the
form

\begin{equation}
\chi (\mathbf{b},s)=P(\mathbf{b})N(\sqrt{s}),
\end{equation}%
where $P(\mathbf{b})$ is the probability distribution of the scattering
center (e.g., partons) in the transverse plane and $N(x)$ is the number of
partons which can interact for a given $\sqrt{s}$. Suppose that $P(\mathbf{b}%
)$ is a two-dimensional Gaussian distribution of width $R,$ 
\begin{equation}
P(\mathbf{b})=\frac{1}{\left( \pi R\right) ^{2}}e^{-\frac{b^{2}}{R^{2}}}
\end{equation}%
and that, for large $\sqrt{s}$, the number of partons increase with $\sqrt{s}
$ as 
\begin{equation}
N(x)\propto \sqrt{s}^{\alpha }.
\end{equation}%
Then, we have 
\begin{equation}
b_{1/2}^{2}\left( \sqrt{s}\right) =\alpha R^{2}\ln \sqrt{s}+Const.
\end{equation}%
In such a situation, the reaction cross section increases as function of the
incident energy for very large $\sqrt{s}$ as 
\begin{equation}
\sigma _{r}\simeq \alpha \pi R^{2}\ln \sqrt{s}.  \label{sig1}
\end{equation}

If the edge of the distribution has an exponential tail instead of a
Gaussian distribution, a similar argument will show that the cross section
would increase as 
\begin{equation}
\sigma _{r}\simeq Const\times \left( \ln \sqrt{s}\right) ^{2}.  \label{sig2}
\end{equation}

The important point of this simple argument is that the rate of increase is
related to the diffuseness $R$ of the probability distribution of the
scattering center. The more diffuse the surface thickness is, the more
quickly the reaction cross section increases as function of the incident
energy. The possibility of such a mechanism, not only in the proton-proton
but also in nucleus-nucleus cross section, was suggested many years ago~\cite%
{Kodama,Kodama1}.

In this paper we explore the idea of~\cite{Kodama,Kodama1} in the language
of the QCD gluon saturation mechanism for the proton-nucleus reaction. We
show that, if the gluon distribution becomes saturated at some energy scale
inside the nuclear surface region, then the reaction cross section of
proton-nucleus collisions starts to increase very quickly and eventually
overcomes the values estimated by the usual Glauber type of calculation.

Applying a simple effective dipole model for the reaction mechanism, we find
that such an energy scale is of the order of $10^{17}-10^{18}eV.$ Above this
energy scale, the behavior of the proton-nucleus cross section begins to
change. We suggest that such an energy dependence of the proton-nucleus
cross section may be observed in terms of the quantity called $\left\langle
X_{\max }\right\rangle $ in the air showers of ultra-high-energy cosmic
rays. Using a very simple toy model estimate of $\left\langle X_{\max
}\right\rangle $, we show that our calculated values of the proton-nucleus
reaction cross section are consistent with the recently observed $%
\left\langle X_{\max }\right\rangle $ by the Pierre Auger Observatory
experiments~\cite{Auger_Xmax} for incident protons at ultra-high energies.

\section{Effective Dipole Model for the Proton-Proton Cross Section}

To explore the above idea, we need the total cross section of proton-proton
as function of incident energy which permit to extrapolate to the ultra-high
energy region.  For this purpose, we introduce an extremely simplified
version of the dipole saturation model for the purpose of illustrating our
idea here. The original dipole model was first introduced by Mueller~\cite%
{Mueller} and extended to the impact parameter representation by Kowalski
and Teaney~\cite{Kowalski}. Recently this model was applied to fit
diffractive structure functions in electron-Nucleus collisions~\cite%
{KowalskieA,KowalskieA2,Bartels}. An application to proton-proton processes
can be found in~\cite{Bartels1}, where the incident proton is represented by
dipoles associated with the valence quarks, having a probability
distribution for different sizes specified by the proton wave function.

In this work, for simplicity, we assume that the proton is described by an
effective dipole of a given size $R_{D}$, where we fix the dipole size $%
R_{D} $ by the average dipole radius

\begin{equation}
R_{D}^{2}=\int d^{2}r_{T}r_{T}^{2}|\Psi _{p}(r_{T})|^{2},  \nonumber
\end{equation}%
where $\Psi _{p}(r_{T})$ is the proton dipole wave function \cite{Erasmo} in
the transverse plane.

In this case, the gluon saturation model leads to a reaction cross section,

\begin{equation}
\sigma _{r}\left( \sqrt{s}\right) =2\pi \int_{0}^{\infty }bdb\ \left(
1-e^{-2\chi }\right) ,  \label{sigma_inel}
\end{equation}%
where the eikonal for the proton-proton reaction can be written as 
\begin{equation}
\chi (\mathbf{b},s)=\frac{\pi ^{2}}{N_{c}}R_{D}^{2}\alpha
_{s}(Q^{2})xg(x,Q^{2})T_{p}(\mathbf{b}).
\end{equation}%
In this expression, the quantity $xg(x,Q^{2})$ is the parton distribution
function (PDF), and represents the gluon density $x-$distribution in the
target at a scale $Q^{2}$, and $\alpha _{s}(Q^{2})$ is the strong coupling
defined at this scale. Following the steps of~\cite{Bartels} we use

\begin{equation}
Q^{2}=\frac{C}{R_{D}^{2}}+Q_{0}^{2},  \label{defscale}
\end{equation}

\begin{equation}
x= \frac{Q^{2}}{Q^{2} + s},
\end{equation}

\begin{equation}
\alpha_{s}(Q^{2})=\frac{4\pi}{\left( 11-\frac{2}{3}N_{f}\right) \log\left( 
\frac{Q^{2}}{\Lambda_{QCD}^{2}}\right) },
\end{equation}
with $N_{f}=3$ and $\Lambda_{QCD}=0.192~GeV$.

For the profile function $T_{p}(\mathbf{b})$ of a proton, we consider the
two different cases described in below.

\subsection{Gaussian profile function}

Suppose that the transverse probability distribution of gluons (profile
function) inside the proton is of a Gaussian type: 
\begin{equation}
T_{p}(\mathbf{b})=\frac{1}{\pi R_{p}^{2}}\exp {\left( \frac{-b^{2}}{R_{p}^{2}%
}\right) .}  \label{Gauss}
\end{equation}%
Writing Eq.(\ref{sigma_inel}) as%
\begin{eqnarray}
&&\sigma _{r}\left( \sqrt{s}\right) =2\pi R_{p}^{2}\int_{0}^{\infty }xdx\
\left( 1-e^{-ae^{-x^{2}}}\right)   \nonumber \\
&=&\pi R_{p}^{2}\int_{0}^{1}\frac{du}{u}\ \left( 1-e^{-au}\right)   \nonumber
\\
&=&\pi R_{p}^{2}\left\{ \gamma _{E}-ChI\left( a\right) +ShI(a)+\ln \left(
a\right) \right\}   \nonumber \\
&\equiv &\pi R_{p}^{2}F(a),  \label{sig_analytic}
\end{eqnarray}%
where 
\begin{equation}
a=\frac{\pi }{N_{c}}\frac{R_{D}^{2}}{R_{p}^{2}}\alpha _{s}(Q^{2})xg(x,Q^{2}),
\label{a}
\end{equation}%
$\gamma _{E}$ is the Euler constant, $ChI\left( a\right) $ and $ShI(a)$ are
hyperbolic-cosine and hyperbolic-sine integral functions. The function%
\begin{equation}
F(a)=\left\{ \gamma _{E}-ChI\left( a\right) +ShI(a)+\ln \left( a\right)
\right\} 
\end{equation}%
is a monotonic increasing function of $a.$ From these expressions, we have $%
a\propto \sqrt{s}^{\alpha }$ and $F\left( a\right) \simeq \ln \left(
a\right) $ for very large $\sqrt{s},$ so that the reaction cross section
behaves at large incident energies as%
\begin{equation}
\sigma _{r}\simeq \alpha \ \pi R_{p}^{2}\ln \left( \sqrt{s}\right) +Const.
\end{equation}%
as expected from the simple argument presented in the Introduction.

\subsection{Profile function with exponential tail}

Another important possibility is that the spatial gluon distribution has an
exponential tail instead of a Gaussian, such as in a Woods-Saxon
distribution. In this case, the asymptotic increase of the cross section as
function of energy is expected as $\log ^{2}\left( \sqrt{s}\right) $ and not 
$\log \left( \sqrt{s}\right) ,$ as discussed in the Introduction. However,
since the Woods-Saxon distribution has 2 parameters (half-density radius $%
R_{1/2}$ and surface thickness $d$), we use the following profile function
which has exponential tail, for simplicity. 
\begin{equation}
T_{p}\left( \vec{b}\right) =\frac{1}{4\pi C_{T}\bar{R}_{p}^{2}}\frac{1}{%
\cosh \left( \frac{b}{\bar{R}_{p}}\right) },  \label{cosh}
\end{equation}%
where $C_{T}$ is the Catalan number, 
\begin{equation}
C_{T}=\sum_{i=0}^{\infty }\frac{\left( -1\right) ^{i}}{\left( 2i+1\right) !}%
\simeq 0.9160.
\end{equation}%
The width parameter $\bar{R}_{p}$ is related to the mean-square of the
impact parameter $\left\langle b^{2}\right\rangle $ as%
\begin{equation}
\left\langle b^{2}\right\rangle =\nu \bar{R}_{p}^{2}.
\end{equation}%
and consequently to the radius parameter of the Gaussian case as $\bar{R}%
_{p}=\sqrt{\nu }R_{p},$ where 
\begin{equation}
\nu =\frac{\int_{0}^{\infty }x^{3}/\cosh x\ dx}{\int_{0}^{\infty }x/\cosh x\
dx}\ \simeq 6.478.
\end{equation}%
Then: 
\begin{eqnarray}
&&\sigma _{r}\left( \sqrt{s}\right) =2\pi \bar{R}_{p}^{2}\int_{0}^{\infty
}xdx\ \left( 1-e^{-a\frac{R_{D}^{2}}{4C_{T}\bar{R}_{p}^{2}\cosh \left(
x\right) }}\right)  \nonumber \\
&=&\pi R_{p}^{2}\times 2\nu I\left( a\right) ,  \label{I(a)}
\end{eqnarray}%
where we defined the integral 
\begin{equation}
I\left( a\right) =\int_{0}^{\infty }xdx\ \left( 1-e^{-\frac{a}{4\nu C_{T}}%
/\cosh \left( x\right) }\right) ,
\end{equation}%
where $a$ is given in Eq.(\ref{a}).

We can examine the asymptotic form of this integral for large $a$ as
follows. Changing variables, $t=\alpha /\cosh x$ with $\alpha =a/4\nu C_{T},$
the integral above can be rewritten as 
\begin{equation}
I\left( \alpha \right) =\int_{0}^{\alpha }\frac{dt}{t}\frac{\left(
1-e^{-t}\right) }{\sqrt{1-\left( \frac{t}{\alpha }\right) ^{2}}}\ln \left( 
\frac{\alpha }{t}+\sqrt{\left( \frac{\alpha }{t}\right) ^{2}-1}\right) .
\end{equation}%
We then separate the integral into two parts: 
\begin{equation}
I\left( \alpha \right) =\left( \int_{0}^{\beta }+\int_{\beta }^{\alpha }\
\right) \frac{dt}{t}\frac{\left( 1-e^{-t}\right) }{\sqrt{1-\left( \frac{t}{%
\alpha }\right) ^{2}}}\ln \left( \frac{\alpha }{t}+\sqrt{\left( \frac{\alpha 
}{t}\right) ^{2}-1}\right) ,
\end{equation}%
where $\beta $ is some finite constant but sufficiently larger than unity so
that $e^{-\beta }\ll 1$. For large $\alpha \gg \beta ,$ the dominant
asymptotic contribution comes from the second integral. In this integral,
the second term $e^{-t}$ may be neglected when compared to unity, and we have%
\begin{equation}
\lim_{\alpha \gg \beta \gg 1}I\left( \alpha \right) \rightarrow \int_{\beta
}^{\alpha }\frac{dt}{t}\ln \left( \frac{2\alpha }{t}\right) =\frac{1}{2}%
\left\{ \ln ^{2}\left( \alpha ^{2}\right) -\ln ^{2}\left( \alpha \beta
\right) \right\} .
\end{equation}%
Therefore, the total reaction cross section at large $\sqrt{s}$ behaves as%
\begin{equation}
\sigma _{r}\simeq 2\nu \pi R_{p}^{2}\ln ^{2}\left( \sqrt{s}\right) +O\left(
\ln \sqrt{s}\right) ,
\end{equation}%
as discussed above.

In Fig.1, we show the behavior of the functions $F\left( a\right) $ and $%
2\nu I\left( a\right) $. Notice that for $a<0.5$ the two functions almost
coincide, but for $a>2$ they behave as $\sim \ln \left( a\right) $ and $\sim
\ln \left( a\right) ^{2},$ respectively.

\subsection{Fits to the proton-proton cross section}

Naturally, our simple effective dipole description will not work well for
the low-energy region. However, the objective of the present work is to show
the effect of possible gluon saturation inside the nuclear surface region at
high energies, we just readjust slightly the parameters determined in~\cite%
{KowalskieA,KowalskieA2} to fit the energy dependence of the proton-proton
reaction cross section~\cite%
{Avila:2002bp,Abe:1993xy,Amos:1991bp,Honda:1992kv,Baltrusaitis:1984ka} only
for $\sqrt{s}>100\ GeV\ $. Note that the cross sections for $\sqrt{s}\simeq
40TeV$ are from the cosmic ray data extracted from the proton-light nuclei
interactions. We can obtain reasonable fits using both Gaussian and
hyperbolic-secant profile functions at higher energies, as seen in Fig.2.
Circles are experimental data, crosses represent the result of the Gaussian
profile function, and open squares are the result of the $1/\cosh $ profile
function. In this energy region, both curves are similar but, as expected,
the cross section for the $1/\cosh $ profile function shows a more rapid
increase in energy at higher energies than the Gaussian profile case. For
these calculations we have used the PDF $xg(x,Q^{2})$ from the GRV98
collaboration~\cite{GRV98}.

\section{Proton-Nucleus cross section}

\subsection{Independent Nucleon Glauber Picture (INGP)}

For a high energy proton-nucleus collision, we may calculate the reaction
cross section $\sigma {}_{p+A}$ as a superposition of independent
nucleon-nucleon collisions in the Glauber approach. Hereafter, this picture
is referred to as INGP. In this picture, we have the well-known formula 
\begin{equation}
\sigma {}_{p+A}=\int d^{2}\mathbf{b~}\left( 1-e^{-AT_{N}\left( \mathbf{\vec{b%
}}\right) \sigma _{pp}\left( \sqrt{s}\right) }\right) ,  \label{Glauber1}
\end{equation}%
where $\sigma _{pp}\left( \sqrt{s}\right) $ is the total cross section of
the proton-proton collision, and $T_{N}\left( \mathbf{b}\right)
=\int_{-\infty }^{\infty }dz\mathbf{\ }P_{A}(\mathbf{b,}z)$ is the
transverse probability distribution function of a nucleon inside the target
nucleus.

The derivation of this formula can be found in many places. Here, for later
comparison, we summarize it. First, assume that the partons are confined in
each nucleon of the target nucleus and, for each collisional event, nucleons
are independently distributed inside the target nucleus. In this vision, for
a one collisional event the incident proton \textquotedblleft
sees\textquotedblright\ the target nucleus as a collection of nucleons whose
center-of-mass positions are specified as 
\begin{equation}
\left\{ \mathbf{b}_{1},\mathbf{b}_{2},..,\mathbf{b}_{A}\right\}
\end{equation}%
in the transverse plane. For this event, we have the eikonal expression of
the total reaction cross section%
\begin{equation}
\sigma _{r}\left\{ \sqrt{s},\mathbf{b}_{1},\mathbf{b}_{2},..,\mathbf{b}%
_{A}\right\} =\int d^{2}\mathbf{b~}\left( 1-e^{-\sum_{i=1}^{A}\chi
_{pp}\left( \mathbf{b-b}_{i};\sqrt{s}\right) }\right) ,  \label{Glauber}
\end{equation}%
where $\chi _{pp}\left( \mathbf{b};\sqrt{s}\right) $ is the eikonal for
proton-proton collisions of impact parameter $\mathbf{b}$.

To calculate the proton-nucleus reaction cross section, we should take the
average over all events. Assuming that the target nucleus is an ensemble of
independent nucleons whose single particle distribution is $P_{A}\left( \vec{%
r}\right) ,$ we can write the average over events as 
\begin{eqnarray}
\left\langle \sigma {}_{p+A}\right\rangle &=&\int d^{3}\vec{r}_{1}\cdots
\int d^{3}\vec{r}_{A}\prod P_{A}\left( \vec{r}_{i}\right) \sigma _{r}\left\{ 
\sqrt{s},\mathbf{b}_{1},\mathbf{b}_{2},..,\mathbf{b}_{A}\right\}  \nonumber
\\
&=&\int d^{2}\mathbf{b~}\left( 1-F(\mathbf{b,}\sqrt{s})^{A}\right) ,
\end{eqnarray}%
where 
\begin{eqnarray}
F(\mathbf{b,}\sqrt{s}) &=&\int d^{3}\mathbf{\vec{r}}^{\prime }\mathbf{\ }%
P_{A}(\mathbf{\vec{r}}^{\prime })e^{-\chi _{pp}\left( \mathbf{b-b}^{\prime };%
\sqrt{s}\right) }  \nonumber \\
&=&1-\int d^{3}\mathbf{\vec{r}}^{\prime }\mathbf{\ }P_{A}(\mathbf{\vec{r}}%
^{\prime })\left( 1-e^{-\chi _{pp}\left( \mathbf{b-b}^{\prime };\sqrt{s}%
\right) }\right) .  \label{F(b)}
\end{eqnarray}%
When $P_{A}(\mathbf{\vec{r}})$ is a slowly varying function compared to the
nucleon size, we can write 
\begin{eqnarray}
F(\mathbf{b,}\sqrt{s}) &=&1-\int d^{2}\mathbf{\vec{b}}^{\prime }\int dz%
\mathbf{\ }P_{A}(\mathbf{\vec{b}}+\mathbf{\vec{b}}_{p},z)\left( 1-e^{-\chi
_{pp}\left( \mathbf{b}_{p};\sqrt{s}\right) }\right)  \nonumber \\
&\simeq &1-\int dz\mathbf{\ }P_{A}(\mathbf{\vec{b}},z)\int d^{2}\mathbf{\vec{%
b}}^{\prime }\left( 1-e^{-\chi _{pp}\left( \mathbf{b}_{p};\sqrt{s}\right)
}\right) .
\end{eqnarray}%
We identify the term 
\begin{equation}
\int d^{2}\mathbf{\vec{b}}^{\prime }\left( 1-e^{-\chi _{pp}\left( \mathbf{b}%
_{p};\sqrt{s}\right) }\right) =\sigma _{pp}\left( \sqrt{s}\right) ,
\end{equation}%
so that 
\begin{equation}
F(\mathbf{b,}\sqrt{s})\simeq 1-\sigma _{pp}\left( \sqrt{s}\right)
T_{z}\left( \mathbf{\vec{b}}\right) ,  \label{F_non_ikonal}
\end{equation}%
where 
\begin{equation}
T_{N}\left( \mathbf{\vec{b}}\right) =\int_{-\infty }^{\infty }dz\mathbf{\ }P(%
\mathbf{b,}z)
\end{equation}%
is the transverse distribution function of a nucleon inside the nucleus.

Eq.(\ref{F_non_ikonal}) becomes negative for $\sigma _{pp}T_{z}>1$ while in
Eq.(\ref{F(b)}) it is defined as positive definite. Considering the
shadowing effect, we should replace Eq.(\ref{F_non_ikonal}) by an
eikonalized expression, 
\begin{equation}
F\rightarrow e^{-T_{z}\left( \mathbf{\vec{b}}\right) \sigma _{pp}\left( 
\sqrt{s}\right) }.
\end{equation}%
From this, we have 
\begin{equation}
\left\langle \sigma {}_{p+A}\right\rangle =\int d^{2}\mathbf{b~}\left(
1-e^{-AT_{N}\left( \mathbf{\vec{b}}\right) \sigma _{pp}\left( \sqrt{s}%
\right) }\right) ,  \label{Sig_Gauber}
\end{equation}%
which is nothing but Eq.(\ref{Glauber1}).

Since 
\begin{equation}
\sigma _{pp}\left( \sqrt{s}\right) \simeq \ln \left( \sqrt{s}\right) ,\ (or\
\ln ^{2}\left( \sqrt{s}\right) ),
\end{equation}%
we conclude that the INGP leads to a very weak energy dependence of the
cross section%
\begin{equation}
\left\langle \sigma {}_{p+A}\right\rangle \sim \ln \ln \left( \sqrt{s}%
\right)   \label{loglog}
\end{equation}%
for large $\sqrt{s},$ and Eq.(\ref{Sig_Gauber}) gives essentially the
geometric cross section of order $\pi R_{N}^{2}.$

We must remind that, in\ Eq.(\ref{Sig_Gauber}) the effects of finite size of
nucleon and of the fluctuation were not included.

\subsection{Gluon saturation in the nuclear surface (GSNS)}

In contrast to the approach above, we may consider the proton-nucleus
collision process in terms of the gluon saturation inside the whole nucleus.
When we go to sufficiently large energies, gluons of bounded nucleons inside
a nucleus should start to superimpose and eventually fill up the nucleus as
a whole. In this regime, we should then use the dipole model with gluon
saturation inside the nucleus to calculate the total cross section for
proton-nucleus collision. Hereafter, such a scenario is referred to as GSNS.

The proton nucleus cross section is then 
\begin{equation}
\sigma _{pA}\left( \sqrt{s}\right) =2\pi \int_{0}^{\infty }bdb\ \left(
1-e^{-2\chi }\right) ,
\end{equation}%
where the eikonal for the proton-nucleus reaction can be written as 
\begin{equation}
\chi (\mathbf{b},s)=\frac{\pi ^{2}}{N_{c}}R_{D}^{2}\alpha
_{s}(Q^{2})xg(x,Q^{2})T_{N}(\mathbf{b}).
\end{equation}%
In this regime, as we discussed before the total reaction cross section
increases as $\ln \left( \sqrt{s}\right) $ or $\ln \left( s\right) ^{2},$
depending on the form of gluon profile function near the nuclear surface
which is much quicker than $\ln \ln \left( \sqrt{s}\right) $ of the
independent nucleon picture. Therefore, the total cross section for
proton-nucleus is eventually dominated by the gluon saturation process
inside the nucleus.

It is interesting to investigate in what energy scale such phenomena will
happen and what physical parameters determine such a scale. In the
independent nucleon picture, the energy dependence of the total cross
section is very weak and stays more or less at the order of the geometrical
cross section, $\pi R_{N}^{2}.$ Therefore, we conclude that the crossover
energy scale should happen when $F\left( a\right) \simeq 1$ or $2\nu I\left(
a\right) $ $\simeq 1$ from Eqs.(\ref{sig_analytic},\ref{I(a)}). From Fig.1,
we find that this occurs at%
\begin{equation}
a\simeq 1,
\end{equation}%
where $a$ is given in Eq.(\ref{a}). For large $\sqrt{s}$, we may approximate
the PDF as 
\begin{equation}
xg\left( \sqrt{s},Q\right) \simeq \xi \left( Q\right) \ \sqrt{s}^{\eta }.
\end{equation}%
The above condition gives an estimate of the energy scale for the crossover: 
\begin{equation}
\sqrt{s}_{Crossover}\simeq \left[ \frac{\pi }{N_{c}}\frac{R_{D}^{2}}{%
R_{N}^{2}}\alpha _{s}(Q^{2})\xi \left( Q\right) \right] ^{-\frac{1}{\eta }}.
\end{equation}%
From this expression, we conclude that the energy scale for the gluon
saturation scenario inside the target nucleus is determined essentially by
the ratio $R_{N}/R_{D}.$ The smaller the ratio is, the smaller the energy
scale becomes.

\section{Ultra high energy cosmic rays}

To see in practice what is the crossover energy scale for the gluon
saturation inside a nucleus, we compare the calculated total reaction cross
sections using the INGP and the GNSN visions. We take typical air nuclei of
average $\left\langle A\right\rangle =14.5\ $where $R_{N}=1.1A^{1/3}\ fm$.
We also calculate the proton-air nuclei cross sections for the following 3
cases of different nuclear profile functions:

\begin{itemize}
\item Gaussian, Eq.(\ref{Gauss}) substituting $R_{p}$ by $R_{N}.$

\item Hyperbolic secant, Eq.(\ref{cosh}) substituting $R_{p}$ by $R_{N},$

\item z-Integrated Woods-Saxon\bigskip 
\begin{equation}
T_{N}(\mathbf{b})=\frac{1}{Z}\int_{-\infty }^{\infty }dz\frac{1}{1+\exp
\left\{ \left( \sqrt{b^{2}+z^{2}}-R_{N}\right) /\alpha \right\} }
\end{equation}%
with $\alpha =0.5$ fm and normalization factor 
\begin{equation}
Z=\frac{4\pi }{3}R_{N}^{3}\left( 1+\pi ^{2}\left( \frac{\alpha }{R}\right)
^{2}\right) .
\end{equation}
\end{itemize}

In Fig. 3 we compare the energy dependences of proton-air collision cross
sections calculated for various situations: One in the INGP $\left( \cdot
\right) ,$ and other 3 cases of the GNSN with Gaussian nuclear profile $%
\left( +\right) $, GNSN with hyperbolic secant profile $\left( \times
\right) $ and the $z-$integrated Woods-Saxon profile $\left( \square \right) 
$. The INGP gives almost the same cross section for all the profile function
so that only one line is shown.

In these calculations, the parameters of the model ($R_{N},R_{p},C,Q_{0}^{2}$
) are those fitted to the proton-proton cross section using the Gaussian
profile (see Table I). In order to calculate the cross section in the
ultra-high-energy cosmic ray domain $\left( >10^{18}eV\right) ,$ we have to
extrapolate the PDF for small $x$ values. For the Gaussian profile we have $%
Q^{2}=18.8~GeV^{2}$ and, consequently, we fit the PDF as 
\begin{equation}
xg\left( x\right) =6.56x^{-0.271}.  \nonumber \\
\end{equation}%
For the $1/\cosh $ case we have $Q^{2}=17.2~GeV^{2}$ and the corresponding
PDF becomes 
\begin{equation}
xg\left( x\right) =6.55x^{-0.267}.  \nonumber \\
\end{equation}%
Naturally, these extrapolations contain ambiguities. However, as we are
interested in obtaining a schematic description of the proton-proton cross
section, these ambiguities can be absorbed in the final fitting procedure of
parameters of the effective dipole model so that they will not affect our
general conclusion. 

As expected, the GNSN scenario gives a rapid increase of the proton-nucleus
cross section as a function of the incident energy and eventually overcomes
the value of the INGP. It is interesting to note that all of the cross
sections for different profile functions of GNSN cross the INGP estimates at
the energy scale of $10^{17}-10^{18}eV.\ $ This is because, once the proton
cross section is fitted, the values of $a$ defined in Eq.(\ref{a}) are not
much different.

As mentioned before, the use of the $1/\cosh $ profile function for the
gluon distribution in a proton gives a more rapid increase of the p-p cross
section at larger energies compared to the Gaussian profile function. When
we use this type of fit to the proton-proton cross section, it naturally
leads to a larger proton-nucleus cross section for the INGP, as shown in
Fig. 4. However, the proton-nucleus cross section calculated in the GSNS
scenario stays invariant. This is because, once the PDF and effective dipole
parameter are determined, the proton-nucleus cross section in this scenario
depends only on the gluon distribution inside the target nucleus.

One direct consequence of such effects should reflect in the behavior of the
observable $\left\langle X_{\max }\right\rangle ,$ essentially the
normalized depth of the position of maximum luminosity of an air-shower in
the atmosphere. This observable can be affected both by the proper increase
of the p-p cross section at ultra-high energy (as in the case of the 1/cosh
profile) and also by the increase of the p-A cross section, due to the gluon
saturation inside the target nucleus. The first possibility was discussed
recently in \cite{Ulrich}. In our case, the p-p cross section fitted by the
1/cosh profile function is very close to the upper limit used in \cite%
{Ulrich}. However, as shown in Fig. 4, this mechanism is less effective than
the GNSN scenario if the INGP description is applied.

To calculate a realistic value of $\left\langle X_{\max }\right\rangle $ we
need a sophisticated simulation of the air-shower processes~\cite%
{Seneca1,Seneca2,corsika} involving all the exclusive cross sections. Here,
just to get an idea on how the above increase of the cross section affects $%
\left\langle X_{\max }\right\rangle ,$ we apply a simple toy model due to
Heitler~\cite{book} to estimate the deviation of $\left\langle X_{\max
}\right\rangle $ from these calculations which are based on the Glauber
description. Assuming such differences of $\left\langle X_{\max
}\right\rangle $ can be identified with the sum of differences of mean free
paths calculated from the INGP and the GSNS scenario as

\begin{equation}
\Delta X_{max}=\sum_{E_{i}>E_{cross}}\Delta X(E_{i}),
\end{equation}%
where 
\begin{equation}
\Delta X(E)=\frac{\left\langle m_{air}\right\rangle }{\sigma _{_{GS}}\left(
E\right) }-\frac{\left\langle m_{air}\right\rangle }{\sigma _{_{Gl}}(E)},
\end{equation}%
and $\left\langle m_{air}\right\rangle $ is the average nuclear mass of the
air, $\sigma _{_{GS}}$ and $\sigma _{_{Gl}}$ are cross sections calculated
by the GNSN scenario and the INGP, respectively. The summation is done over
the cascading steps while the energy of the leading particle is larger than
the crossing energy $E_{Cross}$ from which the gluon saturation scenario
overcomes the nucleon-nucleon Glauber cross picture. To compare with the
experimental data, we use the above toy model estimate for $\Delta X_{\max }$
subtracting from the SIBYLL collaboration results for proton-air $%
\left\langle X_{\max }^{SIBYLL}\right\rangle $ \cite{Sibyll, Ulrich}:%
\begin{equation}
\left\langle X_{\max }\right\rangle \simeq \left\langle X_{\max
}^{SIBYLL}\right\rangle -\Delta X_{\max }.  \label{Xmax}
\end{equation}

In Fig.5, we show the estimated $\left\langle X_{\max }\right\rangle $
values for three different profile functions ($+$ Gaussian,$\ \square $
1/cosh, $\times $ Woods-Saxson) together with the SIBYLL calculations for
the proton-air and Fe-air simulations (dashed lines) and also the observed
values extracted from the Pierre Auger Observatory experiment (black
circles).

\section{Discussion and Perspectives}

In this paper, we have explored the idea of gluon saturation inside a
nucleus in the high-energy limit and its effect on the proton-nucleus cross
section. We show that if gluon saturation occurs in the surface region of
the target nucleus, the proton-nucleus cross section starts to increase very
rapidly as a function of the incident energy. Such a mechanism should
eventually happen for an ultra-high energy scale, but the question is in
which energy scale such a scenario starts to dominate.

Such energy scale is determined by the form of the distribution of gluons
near the surface area. If we assume that the small $x$ gluon distribution in
the nucleus follows that of the nucleon wave function inside the nucleus,
the energy scale where the gluon saturation scenario starts to dominate the
independent nucleon picture is around $10^{17}-10^{18}$ eV. We note that
different profile functions give more or less the same energy scale once the
proton-proton cross section is well fitted. It is very suggestive that the
gluon saturation scenario inside the surface area seems consistent with the
proton primary interpretation of the observed $\left\langle X_{\max
}\right\rangle $ behavior in the Pierre Auger Laboratory experiment.

As we see, the difference between the two pictures, INGP and GSNS, becomes
very large at high energies. The reason for this is that, while in the INGP
the effect of virtual gluons which bound the nucleon near the surface area
is completely neglected, these gluons become dominant at high energies in
the GSNS scenario. In a simple-minded argument, one might think that such an
effect of nuclear binding must be negligible at high energies, since the
ratio of the binding energy of a nucleon to the incident energy tends to
zero. However, the situation may not be so simple. In the GSNS scenario, the
density of virtual gluons, probably forming a kind of fractal fingers when
penetrating into the vacuum among nucleons similar to the electric discharge
pattern, becomes higher and higher at high energies, and eventually
percolate everywhere even in the nuclear surface region. According to the
color glass condensate picture~\cite{Larry,Larry2,Larry3,Larry4}, such a
scenario should happen at some energy scale, even at the lowest density
region of the nuclear surface.

Naturally, the energy scale depends on the precise form of the geometric
gluon distribution inside the nucleus. If the distribution does not follow
the probability distribution of nucleons but more sharp surface
distribution, then the energy scale may shift to higher energy. In the
example of Fig. 3 and Fig. 4 the surface thickness parameter of the
Woods-Saxon distribution is taken a little bit smaller than the usual value
fitted to the nuclear distribution in nucleus ($d\approx 0.6~fm$) since this
fit only applies for heavy nuclei ($A>40$). If we take $d=0.6~fm,$ the
energy scale is lower by one order of magnitude. Therefore, the energy scale
depends crucially on how the gluons starts to saturate in the nuclear
surface region. Depending on this, the energy scale can be even lower than
estimated here. To find a real energy scale where the gluon saturation
occurs at the nuclear surface, further investigation on high-energy
proton-nucleus or electron-nucleus collisions will be necessary. 

There exist ambiguities in the proton-proton cross section used in this
paper (those of the model and extrapolation of the PDF, as well as those of
experimental data) and these affect the precise value of the energy scale.
However, the general conclusion of the present work does not change, as far
as the energy dependence of the proton-proton cross section is fixed.

\section*{Acknowledgments}

This work has been supported by FAPERJ, CAPES, CNPq and PRONEX. The authors
thank Larry McLerran, C. E. Aguiar and E. Fraga for fruitful discussions and
for encouragement, in particular E. Fraga for careful reading the
manuscript. They also acknowledge vivid discussion with the members of the
group ICE of IF-UFRJ during weekly meetings.

\newpage

\section*{\Large Tables}
\addcontentsline{toc}{section}{\Large Tables}

\begin{table}[ht]
\begin{center}
\begin{tabular}{lllll}
\hline
Profile function &  $R_{D}\left( fm\right) $ & $R_{P}\left( fm\right) $ & $C$ & $Q_{0}\left(
GeV\right) $ \\ 
\hline
Gaussian & $0.602$ & 0.621 & 2.8 & 4.3 \\ 
\hline
1/cosh & 0.61 & 0.58 & 3.0 & 4.0 \\
\hline
\end{tabular}
\end{center}
\caption{Parameters of the effective dipole description for the
proton-proton cross section.}
\label{Table1}
\end{table}

\pagebreak

\section*{\protect\Large List of figures}

\addcontentsline{toc}{section}{List of figures}

\begin{figure}[ht]
\centering\includegraphics[scale=0.3]{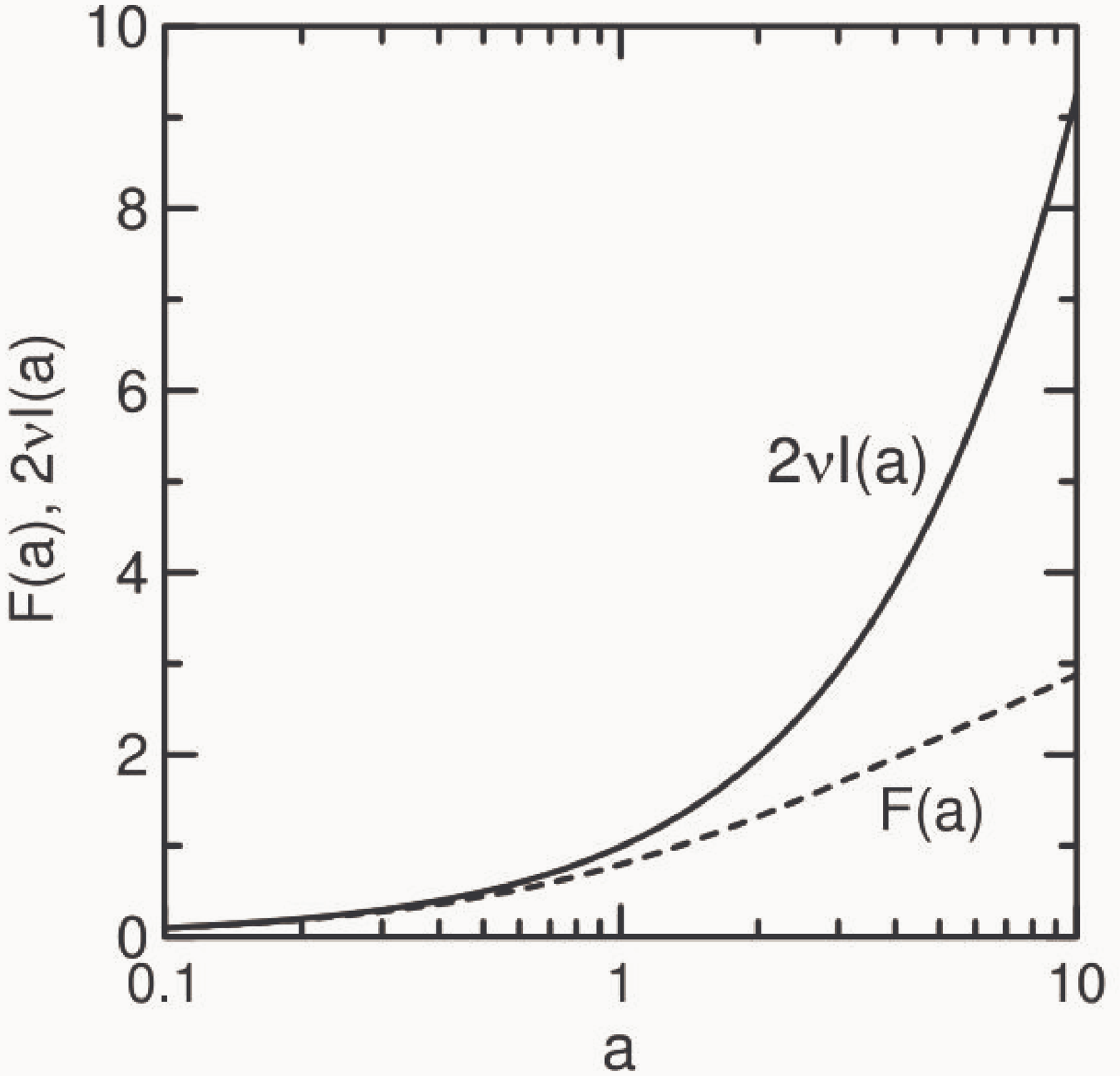} \captiondelim{}
\caption{}
\label{Fig.1}
\end{figure}

\ref{Fig.1}: Comparison of Gaussian and hyperbolic secant profile functions.
The dashed line corresponds to the Gaussian profile function, and the thick
line is that of hyperbolic secant with equivalent radius parameters. \newline

\begin{figure}[ht]
\centering \includegraphics[scale=0.9]{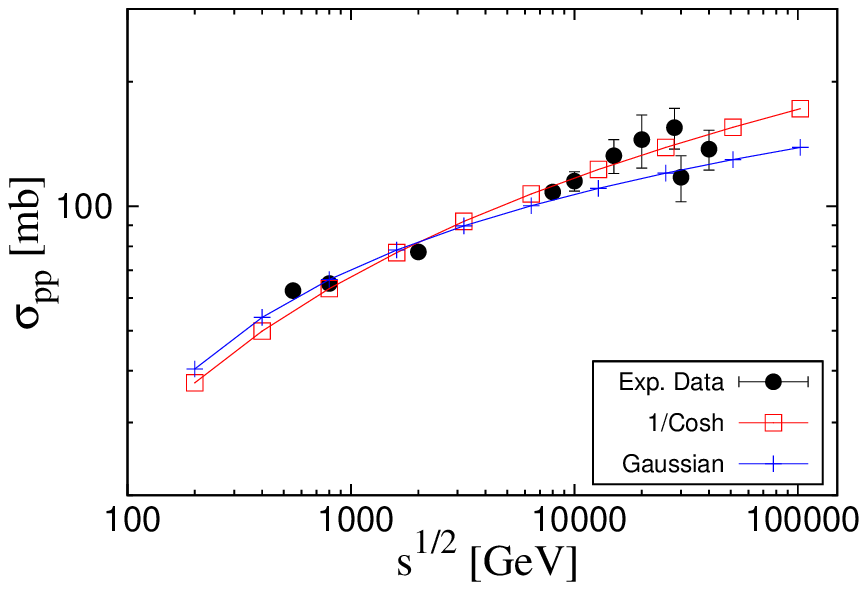} \captiondelim{}
\caption{}
\label{Fig.2}
\end{figure}

\ref{Fig.2}: Fits to proton-proton cross sections. Circles are experimental
data~ \cite%
{Avila:2002bp,Abe:1993xy,Amos:1991bp,Honda:1992kv,Baltrusaitis:1984ka}
crosses are the Gaussian profile function and the open squares are for the
hyperbolic secant function.\newline

\begin{figure}[ht]
\centering \includegraphics[scale=0.9]{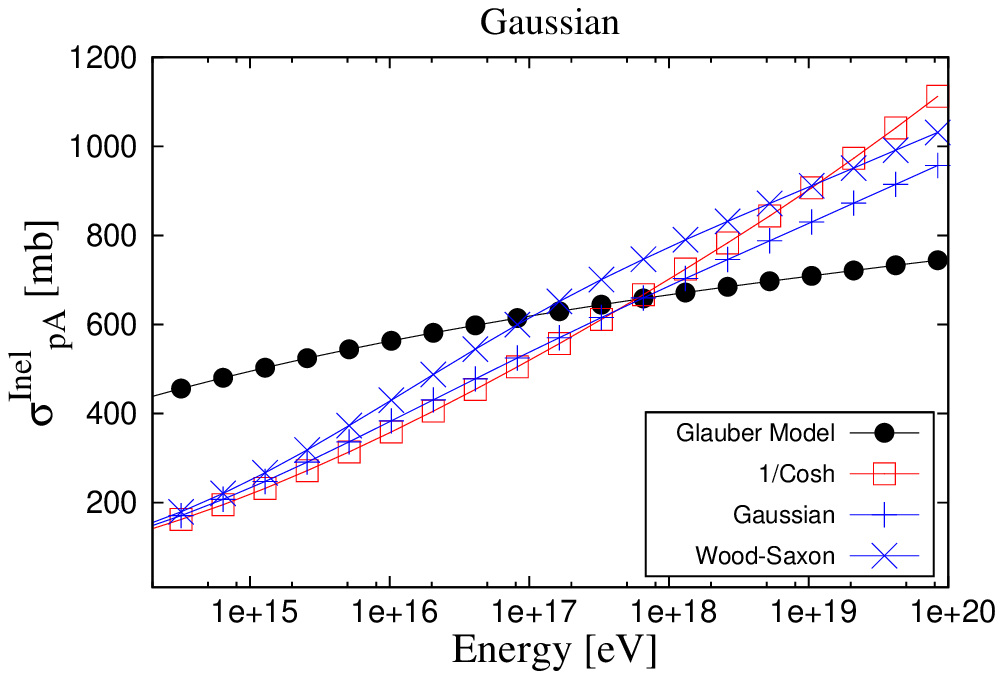} \captiondelim{}
\caption{}
\label{Fig.3}
\end{figure}

\ref{Fig.3}: Proton-Nucleus cross sections. Black circles are for the INGP,
and other three $\left( +,\times,\square\right) $ correspond to the secnario
of GSNS.\newline

\begin{figure}[ht]
\centering \includegraphics[scale=0.9]{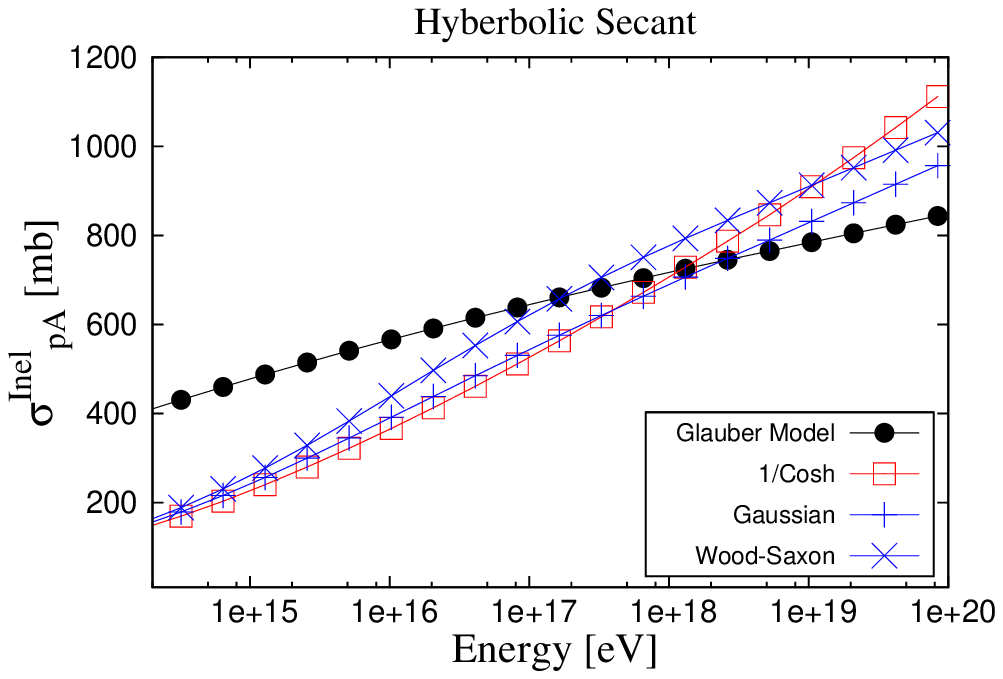} \captiondelim{}
\caption{}
\label{Fig.4}
\end{figure}

\ref{Fig.4}: Proton-Nucleus cross sections using the proton cross section
fit by $1/\cosh$ profile function. Legends are the same as Fig.3.\newline

\begin{figure}[tbp]
\centering \includegraphics[scale=0.5]{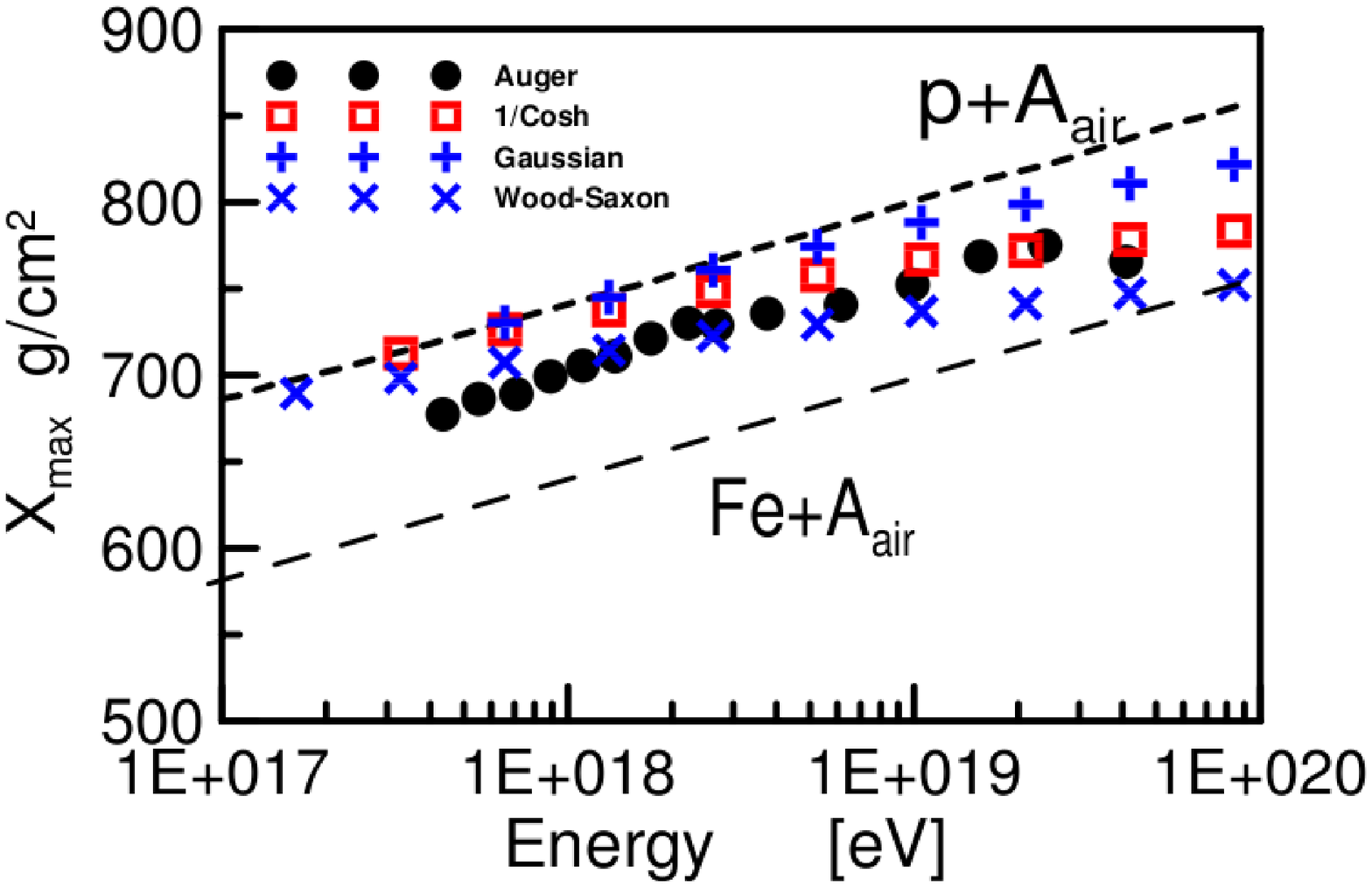} \captiondelim{}
\caption{}
\label{Fig.5}
\end{figure}

\ref{Fig.5}: Estimated $\left\langle X_{\max}\right\rangle$ using the
Heitler model. The dotted (proton-air) and dashed (Fe-air) lines are taken
from SIBYLL collaborations and black circles are the observed values
extracted from the Auger Experiment. Our $\left\langle X_{\max} \right\rangle
$ are calculated for three different profile functions ($+$ Gaussian,$\
\square$ 1/cosh, $\times$ Woods-Saxson) as the deviation from the upper line
according to Eq.(\ref{Xmax}). \newline

\pagebreak

\section*{\protect\Large Figures}

\addcontentsline{toc}{section}{Figures}

\pagebreak

\pagebreak

\pagebreak

\pagebreak

\end{document}